\documentclass[aps,prd,floatfix,twocolumn,superscriptaddress,preprintnumbers,nofootinbib]{revtex4-1}

\usepackage{textcomp} 
\usepackage{slashed}
\usepackage{epsfig,latexsym,cancel,amssymb,amsmath,verbatim,mathrsfs}
\usepackage{color}
\usepackage{graphicx}
\usepackage{enumitem}
\usepackage[colorlinks,citecolor=blue]{hyperref}

\usepackage{slashed}
\allowdisplaybreaks[4]

\begin{document}

\title{Rotational Metric: A Solution to Einstein's Clock-Rate Problem \\and Its Magnetospheric Applications}

\author{Zhen Zhang}
\email{zhangzhen@ihep.ac.cn}
\affiliation{State Key Laboratory of Particle Astrophysics, Institute of High Energy Physics, Chinese Academy of Sciences, 
19B Yuquan Road, Beijing 100049, People's Republic of China}

\author{Rui Zhang}
\email{rui.z@pku.edu.cn}
\affiliation{Theoretical Physics Division, Institute of High Energy Physics, Chinese Academy of Sciences, 
19B Yuquan Road, Beijing 100049, People's Republic of China}
\affiliation{Shanghai Key Laboratory of Deep Space Exploration Technology, Shanghai Institute of Satellite Engineering, 3666 Yuanjiang Road, Shanghai 201109, People’s Republic of China}

\begin{abstract}
\vspace{0mm}
\noindent
{\bf Abstract:} The rotational metric provides an exact solution to Einstein's clock-rate problem in curved spacetime, specifically, whether time flows more slowly at the equator of a compact object such as a neutron star than at its poles. It features a curvature singularity, an event horizon, a potentially evolving ergosphere, a rigidly-rotating normal space, and two stationary limit surfaces. Although derived from the Schwarzschild metric through rotational transformations, it includes an additional ergosphere. Given the equivalence of inertia and gravity, this demonstrates how non-inertial transformations, such as rotational transformations, can introduce new spacetime structures into a gravitational system. In particular, the additional physical degrees of freedom carried by rotational transformations are `eaten' by the gravitational system to form an additional ergosphere. Furthermore, the rotational metric effectively models a rigidly-rotating gravitational system and is applicable for describing rotationally-induced gravitational effects in various rotating magnetospheres.

\vspace{0mm}
\end{abstract}

\maketitle


\section{Introduction}

By Einstein's 1905 article on the special theory of relativity, it was concluded that, under identical conditions, a balance-clock placed at the equator must run more slowly than an identical clock located at one of the poles \citep{Einstein1952}. 
In 1909, Einstein constructed a metric for a uniformly rotating disk \citep{Einstein:1916vd} that contributed to the development of a key idea in general relativity (GR): the relevance of curved geometry in the presence of gravity \citep{Weinberg:1972kfs,Rindler2006book,Carroll2014}.
Notably, this metric includes non-inertial effects, such as the centrifugal (inertial) forces, that arise from rotation. 
By Einstein's equivalence principle, inertia and gravity are identical. 
In GR, gravity distorts Minkowskian spacetime, and so does inertia.
Therefore, the geometry of a uniformly rotating disk is curved; for example, its circumference is shortened owing to this curvature.
Approximately six years later, in 1915, Einstein proposed the full GR field equations and developed the general theory of relativity \citep{Rindler2006book}.
Then, in 1916, Schwarzschild found the first exact solution to the Einstein field equations \citep{Schwarzschild03}, 
which can be used to describe the highly curved spacetime around a compact object such as a neutron star (NS).
Given the rotation of the compact object, observers standing on its surface must confront the clock-rate problem in curved spacetime:
{\it For a spinning compact object, does time pass more slowly at the equator than at the poles?}
In light of the highly curved geometry of spacetime, this problem should be addressed within the framework of GR. 
Conversely, the solution to this problem may have direct applications in everyday life.
For instance, when accurately measuring how time passes at two locations on Earth, we must account for the spacetime effects caused by the Earth's mass and rotation.
In fact, precisely resolving the clock-rate problem in curved spacetime is particularly important for technologies that depend on accurate time measurements, such as the BeiDou Navigation Satellite System and Global Positioning System.

NSs were first discovered in 1967, and it quickly became apparent to astrophysicists that the picture of a rotating magnetic NS existing in an empty space was not realistic \citep{1982RvMP...54....1M}. Early in 1969, Goldreich and Julian argued that a rapidly spinning, highly magnetized NS would create a magnetosphere around the star \citep{Goldreich96}. Charged particles are pulled out from the NS surface, form a magnetospheric plasma, and then become frozen into the NS's magnetic field, resulting in steady and rigid corotation with the star \citep{1982RvMP...54....1M,Goldreich96}. A magnetosphere is generally assumed to be the source of various emissions \citep{doi:10.1146/annurev-astro-081915-023329}, e.g., the well-known fast radio bursts \citep{CHIMEFRB:2020abu,Bochenek:2020zxn}, giant flares \citep{watts2006detection,strohmayer2005discovery,Aptekar_2001,Yi:2023mph}, magnetar X-ray bursts \citep{hxmt2020a,kw2020a,in2020a}, and super flares \citep{Zhang:2022qtd,Xiao:2022quv} observed in magnetized NSs. To date, the corotational effects of this gravitational system have not been fully studied. To better understand the origin and nature of these high-energy phenomena, it is essential to construct or employ a metric that effectively describes the gravitational or non-inertial effects induced by rotation within any corotating magnetosphere.  It is worth noting that, in addition to NSs, magnetospheres exist around the Earth, normal stars, white dwarfs, quark stars, and various other exotic astrophysical objects.

Generally, the Kerr metric can describe rotating gravitational systems \citep{Kerr1963} such as black holes (BHs).
Regrettably, the Kerr metric cannot account for rigid corotation, as the angular velocity varies with radius \citep{GrHer2007book}.
Notably, the standard model of magnetospheres typically involves a space volume that maintains a rigid corotation with a spinning mass \citep{Goldreich96}. 
The rotation in this case differs from that described by the Kerr metric. 
However, it is possible to introduce rigid rotation into the Schwarzschild metric through rotational transformations that carry additional physical degrees of freedom.
In this study, by applying such transformations, we developed a metric
that describes the gravitational system of a rigidly-rotating space volume. 
This metric remains a valid solution to the vacuum Einstein equations.
Interestingly, it can be used to resolve Einstein's clock-rate problem in curved spacetime.
Additionally, the metric includes an angular-momentum parameter that may evolve over time, corresponding to additional physical degrees of freedom.
Given the inertia-gravity (or the gravity of inertia) \citep{Rindler2006book} induced by rotation, it can also be used as an example to demonstrate the non-inertial effects associated with rotational transformations.
We investigated the geometric structures of the metric as well as its gravitational effects and direct applications.
We summarize our findings at the end of this paper.

\section{Rotational Metric and Its Interpretations} 
\label{sec:2} 

The Schwarzschild metric is the unique static and spherically symmetric vacuum solution in GR.
In spherical coordinates $\left(t, r, \theta, \varphi\right)$, it has the form 
\begin{equation}
\label{eq:SchMsph}
\begin{array}{rcl}
\displaystyle
\mathrm{d} s^2 \,=\,\!&-& \left(1\!-2\,\frac{\,M\,}{r}\right)\,\mathrm{d} t^{2}\\[2mm]
&+&\frac{1}{\, \left(1-2\,\frac{\,M\,}{r}\right) \,}\,\mathrm{d} r^2\!+ r^{2}\left(\mathrm{d} \theta^2+\sin^{2}\theta\,\mathrm{d}\varphi^2\right),
\end{array}
\end{equation}
where $M$ is the mass of the gravitational system. The geometrized unit system $G=c=1$ is adopted. 
In general, the Schwarzschild metric can properly describe a static gravitational system.

A gravitational system that rotates rigidly can be effectively described by a stationary metric, which can be expressed in the canonical form~\citep{Rindler2006book},
\begin{equation}
\label{eq:CF}
\begin{array}{rcl}
\displaystyle
\mathrm{d} s_{\omega}^2 \,=\,\!&-&e^{2\psi}\,\left(\mathrm{d} t-\upsilon_{i}\,\mathrm{d} x^{i}\right)^{2}+k_{ij}\,\mathrm{d} x^{i}\mathrm{d} x^{j}, 
\end{array}
\end{equation}
with $i, j=1, 2, 3$, where $t$, $x^{i}$, and $x^{j}$ are the time-like and space-like coordinates, while the functions $\psi$, $\upsilon_{i}$, and $k_{ij}$ are independent of time.

To determine the specific form of the metric given by Eq.\,\eqref{eq:CF} for a rigidly-rotating system, we rewrite the Schwarzchild metric given by Eq.\,\eqref{eq:SchMsph} in cylindrical coordinates:
\begin{equation}
\label{eq:SchMcyl}
\begin{array}{rcl}
\displaystyle
\nonumber
\mathrm{d} s^2 \,=\,\!&-& \left(1\!-2\,\frac{\,M\,}{r}\right)\,\mathrm{d} t^{2}\\[2.5mm]
&+&\bigg[1+\left(\frac{2\frac{M}{r}}{1-2\frac{M}{r}}\right)\frac{\rho^2}{r^2}\bigg]\mathrm{d} \rho^2+2\bigg[\left(\frac{2\frac{M}{r}}{1-2\frac{M}{r}}\right)\frac{z\rho}{r^2}\bigg]\mathrm{d} \rho\,\mathrm{d} z\\[3.5mm]
&+&\bigg[1+\left(\frac{2\frac{M}{r}}{1-2\frac{M}{r}}\right)\frac{z^2}{r^2}\bigg]\mathrm{d} z^2
+\rho^2\mathrm{d} {\varphi^{\prime}}^2,
\end{array}
\end{equation}
where $\rho=\sqrt{r^2-z^2}$ is the cylindrical radius, $z$ is the height over the equatorial plane, and $\varphi^\prime$ is the azimuth angle measured by a static observer at spatial infinity.

For a rotating gravitational system with a constant angular velocity $\omega$, the azimuth angle $\varphi$ is expressed as
\begin{equation}
\label{eq:phi-dphi0}
\varphi=\varphi^{\prime}+ \omega\, t, 
\end{equation}
where the angular coordinate $\varphi$ is measured from the $\left(\rho, z\right)$-plane that rotates about the $z$-axis,
which is employed for deriving the Einstein's metric for a uniformly rotating disk.
However, if $\omega$ is a function of time, i.e., $\omega=\omega(t)$, one sets 
\begin{equation}
\label{eq:phi-dphi}
\mathrm{d} \varphi=\mathrm{d} \varphi^{\prime}+ \omega\,\mathrm{d} t,
\end{equation}
which notably differs from \eqref{eq:phi-dphi0}. From this point on, our derivations and analysis are carried out based on equation~\eqref{eq:phi-dphi} instead of \eqref{eq:phi-dphi0}.
Applying the transformation given by~\eqref{eq:phi-dphi}, one obtains the canonical form of the metric expressed by Eq. \eqref{eq:CF} for a stationary system,
\begin{equation}
\label{eq:RotSchM}
\begin{array}{rcl}
\displaystyle
\mathrm{d} s_{\omega}^2 \!=\!&\!-\!&\!\!\left(\!1\!-\!2\!\frac{\,M\,}{r}\!-\!\rho^2 \omega^2\!\right)\!\bigg[\mathrm{d} t\!+\!\left(\!\frac{\rho^2 \omega}{1\!-\!2\,\frac{\,M\,}{r}\!-\!\rho^2 \omega^2}\right)\mathrm{d} \varphi\bigg]^{2}\\[4mm]
\!&+&\!\bigg[\left(\frac{1\!-\!2\,\frac{\,M\,}{r}}{1\!-\!2\,\frac{\,M\,}{r}-\!\rho^2 \omega^2}\right)\rho^2\bigg]\mathrm{d} \varphi^2\!+\!\mathrm{d}\sigma^{2},
\end{array}
\end{equation}
where $\rho < r_{{\rm LC}}=1/\omega$, and $\mathrm{d}\sigma^{2}$ is expressed as
\begin{equation}
\label{eq:dsigma2}
\begin{array}{rcl}
\displaystyle
\mathrm{d}\sigma^{2}=&&\bigg[1+\left(\frac{2\frac{M}{r}}{1-2\frac{M}{r}}\right)\frac{\rho^2}{r^2}\bigg]\mathrm{d} \rho^2+2\bigg[\left(\frac{2\frac{M}{r}}{1-2\frac{M}{r}}\right)\frac{z\rho}{r^2}\bigg]\mathrm{d} \rho\,\mathrm{d} z\\[4mm]
&&+\bigg[1+\left(\frac{2\frac{M}{r}}{1-2\frac{M}{r}}\right)\frac{z^2}{r^2}\bigg]\mathrm{d} z^2.
\end{array}
\end{equation}
The spatial part of the metric is expressed as
\begin{equation}
\label{eq:dl2}
\begin{array}{rcl}
\displaystyle
\mathrm{d}{l}^{2}\!&\!=\!&\!k_{ij}\,\mathrm{d} x^{i}\mathrm{d} x^{j}
\!=\!\bigg[\left(\frac{1\!-\!2\,\frac{\,M\,}{r}}{1\!-\!2\,\frac{\,M\,}{r}\!-\!\rho^2 \omega^2}\right)\,\rho^2\bigg]\mathrm{d}\varphi^2\!+\!\mathrm{d}\sigma^{2},
\end{array}
\end{equation}
where $r_{{\rm LC}}$ is the light cylinder radius \citep{Goldreich96,1982RvMP...54....1M}, widely used in the physics of NSs. 
We use $x^{\mu}~\left(\mu=0,~1,~2,~3\right)$ for $\left(t,~\rho,~z,~\varphi\right)$. The general form of the stationary metric \eqref{eq:RotSchM} in cylindrical coordinates is 
\begin{equation}
\label{eq:GRotSchM}
\begin{array}{rcl}
\displaystyle
\mathrm{d} s_{\omega}^2 =\!g_{00}\,\mathrm{d} t^2\!+\!g_{0i}\,\mathrm{d} t \,\mathrm{d} x^{i}\!+\!g_{i0}\,\mathrm{d} x^{i} \,\mathrm{d} t\!+\!g_{ij}\,\mathrm{d} x^{i} \mathrm{d} x^{j},
\end{array}
\end{equation}
where $g_{0i}=g_{i0}$ and $g_{ij}=g_{ji}$. The components of the metric are exactly expressed as
\begin{equation}
\label{eq:Gmetric}
\begin{array}{rcl}
\displaystyle
\!\!g_{\mu\nu }\!\!=\!\!\!\left[\!
\begin{matrix}
\!-\!1\!+\!\frac{2 M}{r}\!+\!\omega^2 \rho ^2 \!\!&\!\! 0 \!\!&\!\!0 \!\!&\!\! -\omega\rho ^2\! \\
 \!0 \!\!&\!\! 1\!+\!\frac{2 M \rho ^2}{r^{3} \left(1\!-\!\frac{2 M}{r}\right)} \!\!&\!\! \frac{2 M \rho  z}{ r^{3} \left(1\!-\!\frac{2 M}{r}\right)} \!\!&\!\! 0\! \\
 \!0 \!\!&\!\! \frac{2 M \rho  z}{ r^{3} \left(1\!-\!\frac{2 M}{r}\right)} \!\!&\!\! 1\!+\!\frac{2 M z^2}{r^3 \left(1\!-\!\frac{2 M}{r}\right)} \!\!&\!\! 0 \!\\
\!-\omega\rho ^2  \!\!&\!\! 0 \!\!&\!\! 0 \!\!&\!\! \!\rho ^2\! \\
\end{matrix}
\right]\!,\!
\end{array}
\end{equation}
which imply that the metric is flat at spatial infinity in the direction of the $z$-axis, whereas it is not asymptotically flat in any other direction.
The metric $\mathrm{d} s_{\omega}^2$ has been presented in its full form for a $1\!+\!3$ dimensional spacetime, without being limited to its equatorial plane.

In analogy with the original Schwarzschild metric, the rotational metric $\,\mathrm{d} s_{\omega}^2\,$ has a genuine curvature singularity at $r=0$, which is confirmed by deriving the Kretschmann invariant,
\begin{equation}
\label{eq:Curvature4}
\begin{array}{rcl}
\displaystyle
K=R^{\mu\nu\varrho\sigma}R_{\mu\nu\varrho\sigma}=48\,\frac{M^2}{r^6}, 
\end{array}
\end{equation}
where $R^{\mu\nu\varrho\sigma}$ and $R_{\mu\nu\varrho\sigma}$ are the curvature tensors of types $(4, 0)$ and $(0, 4)$, respectively. 
Note that the metric $\mathrm{d} s_{\omega}^2$ exhibits a similar characteristic near the curvature singularity as the original Schwarzschild metric does before transformations.

For any stationary gravitational system, its angular velocity ${\mathrm\Omega}$ can be generally expressed in cylindrical coordinates as \citep{Rindler2006book,GrHer2007book,Carroll2014}
\begin{equation}
\label{eq:AngularVelocity0}
\nonumber
{\mathrm\Omega}=-\frac{g_{03}}{g_{33}}.
\end{equation}
In our case, the quantity $\omega$ in the metric $\mathrm{d} s_{\omega}^2$ is simply the angular velocity of a corresponding rotating system.
Indeed, one finds
\begin{equation}
\label{eq:AngularVelocity}
{\mathrm\Omega}=\omega,
\end{equation}
which means that the gravitational system associated with the metric $\mathrm{d} s_{\omega}^2$ rigidly rotates at angular velocity $\omega$ with respect to the $z$-axis.
As expected, it can reduce to the original Schwarzschild metric in the zero-rotation limit $\omega\to0$ and to Einstein's metric for a uniformly rotating disk \citep{Rindler2006book} at the zero-mass limit $M\to0$. Hence, the rotational metric is an extension of the Einstein's metric for a uniformly rotating disk, which rigidly corotates with a spinning mass around a fixed axis. It is straightforward to demonstrate that the {\it rotational metric} $\mathrm{d} s_{\omega}^2$ remains a vacuum solution of the Einstein field equations; more interpretations and derivations about the rotational metric are presented in appendices.

Although derived from the Schwarzschild metric through coordinate transformations, 
the rotational metric includes additional effects of inertia-gravity arising from rotation.
Hence, the rotational and Schwarzschild metric can describe two different gravitational systems: a rigidly-rotating system and a static system, respectively. 
When $\omega$ is a constant, the transformation given by Eq. \eqref{eq:phi-dphi} leads to the asymptotic non-flatness of the rotational metric $\mathrm{d} s_{\omega}^2$. The condition of asymptotic flatness is necessary in the unicity theorem for the Kerr solution \citep{Carter:1971zc}. Therefore, for a constant $\omega$, the existence of the rotational metric $\mathrm{d} s_{\omega}^2$ does not contradict the uniqueness of the Kerr metric in an asymptotically-flat stationary spacetime.
In the general case of $\omega\equiv\omega(t)$, the rotational metric $\mathrm{d} s_{\omega}^2$ is a time-evolving metric. 
However, the time evolution of the rotational metric $\mathrm{d} s_{\omega}^2$ may arise from the energy conversion of the rotational energy of the gravitational system into other forms of energy through non-gravitational interactions or quantum-mechanical processes, which is far beyond the scope of this study.

\section{Horizon Structures and Non-Inertial Transformations}  
\label{sec:3} 

According to GR, gravity distorts flat spacetime. 
Given that inertia and gravity are equivalent, non-inertial effects, such as rotation, may have influence on the spacetime structures of a BH.
In particular, for a BH described by the rotational metric, such influence should be carefully investigated.
In general, the event horizon of a BH is a null surface defined by $f=f\left(\rho,z\right)=constant$. 
The gradient $\partial_{\mu}f$ is normal to the null surface; it is a null vector. By definition, one obtains a null surface equation, i.e., $\partial_{\mu}f\,\partial^{\mu}f=0$.
From the rotational metric, we obtain the components of its inverse:  
\begin{equation}
\label{eq:IGmetric}
g^{\mu\nu}\!=\!\left[
\begin{matrix}
-\frac{1}{1\!-\!\frac{2 M}{r}} & 0 &0 & -\frac{\omega}{1\!-\!\frac{2 M}{r}} \\
 0 & 1\!-\!\frac{2 M \rho ^2}{r^{3}} & -\!\frac{2 M \rho  z}{ r^{3} } & 0 \\
 0 & -\!\frac{2 M \rho  z}{ r^{3} } & 1\!-\!\frac{2 M z ^2}{r^{3}} & 0 \\
-\frac{\omega}{1\!-\!\frac{2 M}{r}}  & 0 & 0 & \frac{1\!-\!\frac{2 M}{r}\!-\!\omega^2\!\rho ^2}{1\!-\!\frac{2 M}{r}}\frac{1}{\rho^2} \\
\end{matrix}
\right].
\end{equation}
Substituting these components into the null surface equation yields
\begin{equation}
\label{eq:ehorizon}
\left(\frac{\partial f}{\partial \rho}\right)^2+\left(\frac{\partial f}{\partial z}\right)^2-2\frac{M}{r}\left(\frac{\rho}{r}\frac{\partial f}{\partial \rho}+\frac{z}{r}\frac{\partial f}{\partial z}\right)^2=0,
\end{equation}
which is symmetric in $\rho$ and $z$ and hence also $f$'s expression.
When $f=f\left(r\right)$, Eq.~\eqref{eq:ehorizon} can be rewritten as 
\begin{equation}
\label{eq:ehorizon2}
\left(1-\frac{2M}{r}\right)\left(\frac{\partial f}{\partial r}\right)^2=0.
\end{equation}
Thus, the event horizon occurs at radius $r_{\rm EH}=2M$, where $r_{\rm EH}$ is the same as the Schwarzschild radius.
Note that the Kretschmann invariant remains finite there.

Besides, the BH system has two stationary limit surfaces (SLSs), which can be obtained by solving the equation
\begin{equation}
\label{eq:limitSurf}
g_{00}=-\left(1\!-\!\frac{2 M}{r}\!-\!\omega^2 \rho ^2\right)=0.
\end{equation}
For each of them, one always has $1-\left(\frac{\rho}{r_{{\rm LC}}}\right)^2>0$, i.e., $\omega\rho<1$. Therefore,
\begin{equation}
\label{eq:limitSurf2}
r= \frac{2M}{1-\left(\frac{\rho}{r_{{\rm LC}}}\right)^2}> r_{\rm EH},
\end{equation}
which clearly indicates that the two SLSs are outside the event horizon.
Two SLSs are approximately found at $r\sim r_{\rm EH}$ and $r\sim r_{{\rm LC}}$, which will be referred to as the inner and outer SLSs, respectively.
In spherical coordinates, $\rho=r \,\mathrm{sin}\,\theta$ and $z=r \,\mathrm{cos}\,\theta$. 
For $\omega\rho<1$, Eq.~\eqref{eq:limitSurf} becomes a univariate cubic equation in $r$, yielding one negative solution and two positive solutions. Clearly, the negative solution is physically unfeasible, whereas the positive solutions represent the two SLSs, which can be parametrized using spherical coordinates $\left(r,\theta\right)$ as
\begin{equation}
\label{eq:rlimit}
\begin{array}{rcl}
\displaystyle
r^{\pm}_{\theta}\!\!=\!\!&\!\!&\!\!\!\!M\,\bigg[\frac{\frac{1\pm i\sqrt{3}}{2} \sqrt[3]{\!1\!+\!i\! \sqrt{\frac{1}{27 \chi_{\theta}^2}\!-\!1}}}{\chi_{\theta}^{2/3}}
\!+\!\frac{\frac{\!1\!-\!\left(\!\pm\!\right)\! i \! \sqrt{3}}{2} \sqrt[3]{\!1\!-\!i \! \sqrt{\frac{1}{27 \chi_{\theta}^2}-1}}}{ \chi_{\theta}^{2/3}}\bigg],
\end{array}
\end{equation}
where $``-"$ and $``+"$ represent the inner and outer SLSs, respectively. 
Here, the quantity $\chi_{\theta}$ can be further expressed as 
\begin{equation}
\label{eq:ttheta}
\chi_{\theta}=\frac{1}{~2~}\frac{r_{\rm EH}}{r_{{\rm LC}}}\,\mathrm{sin}\,\theta.
\end{equation}
Note that the two terms in Eq.~\eqref{eq:rlimit} are complex conjugates of each other. Thus, both $r^{-}_{\theta}$ and $r^{+}_{\theta}$ are real functions.
In the range of $0<\chi_{\theta}\leq\frac{1}{3\sqrt{3}}$, both are positive.
For $\theta=\pi/2$, when $\chi_{\theta}=\frac{1}{~2~}\frac{r_{\rm EH}}{r_{{\rm LC}}}=\frac{1}{3\sqrt{3}}$, 
the inner and outer SLSs begin to touch each other at the point $\left(\rho,\,z\right)=\left(3/2\, r_{\rm EH},\,0\right)$. 
It means that $\omega M=\frac{1}{~2~}\frac{r_{\rm EH}}{r_{{\rm LC}}}$ has an upper limit of $\frac{1}{3\sqrt{3}}$.
Let us define $\Delta=r^{-}_{\theta=\frac{\pi}{2}}-r_{\rm EH}$.
Then, $0<\Delta/r_{\rm EH}\leq50\%$, and equality holds if and only if $\omega M$ takes the upper limit.

A massive test particle travels along a time-like path.
If the movement of the test particle is constrained to $\rho=constant$ and $z=constant$,
the spacetime experienced by the particle should be time-like. Consequently, one has $\mathrm{d} s_{\omega}^2 \,=\,\!g_{00}\,\mathrm{d} t^2+2 \,g_{03}\,\mathrm{d} t \,\mathrm{d} \varphi+g_{33}\,\mathrm{d} \varphi^2<0$.
Let us denote $\omega_{\mathrm p}=\frac{\mathrm{d} \varphi}{\mathrm{d} t}$ as the angular velocity of the test particle.
From this point on, we assume $\omega>0$ without loss of generality.
Therefore, we have
\begin{equation}
\label{eq:wp}
g_{00}+2 \,g_{03}\,\omega_{\mathrm p}+g_{33}\,\omega_{\mathrm p}^2<0.
\end{equation}
This can be equivalently expressed as
\begin{equation}
\label{eq:wpm}
\omega_{\mathrm -}\!=\!\omega-\frac{1}{\rho}\sqrt{\!1\!-\!\frac{2M}{r}\!}\!<\!\omega_{\mathrm p}\!<\!\omega_{\mathrm +}\!=\!\omega\!+\!\frac{1}{\rho}\sqrt{\!1\!-\!\frac{2M}{r}}~,
\end{equation}
which imposes constraints on the angular velocity of the test particle. 
In the region defined by $r_{\rm EH}<r<r^{-}_{\theta}$, we always have $\omega_{\mathrm -}>0$. 
Thus, this region is where the time-like path of the test particle is inevitably dragged along with the BH's rotation.
In GR, this region is known as the ergosphere.
We have rigorously demonstrated the existence of an ergosphere between the event horizon and the inner SLS.
Note that this result remains valid even if the dynamical parameter $\omega$ is time-dependent. 
Therefore, the evolution of the ergosphere over time can be demonstrated based on the rotational metric.

\begin{figure}
\centering
\resizebox{0.8\columnwidth}{!}{
\includegraphics{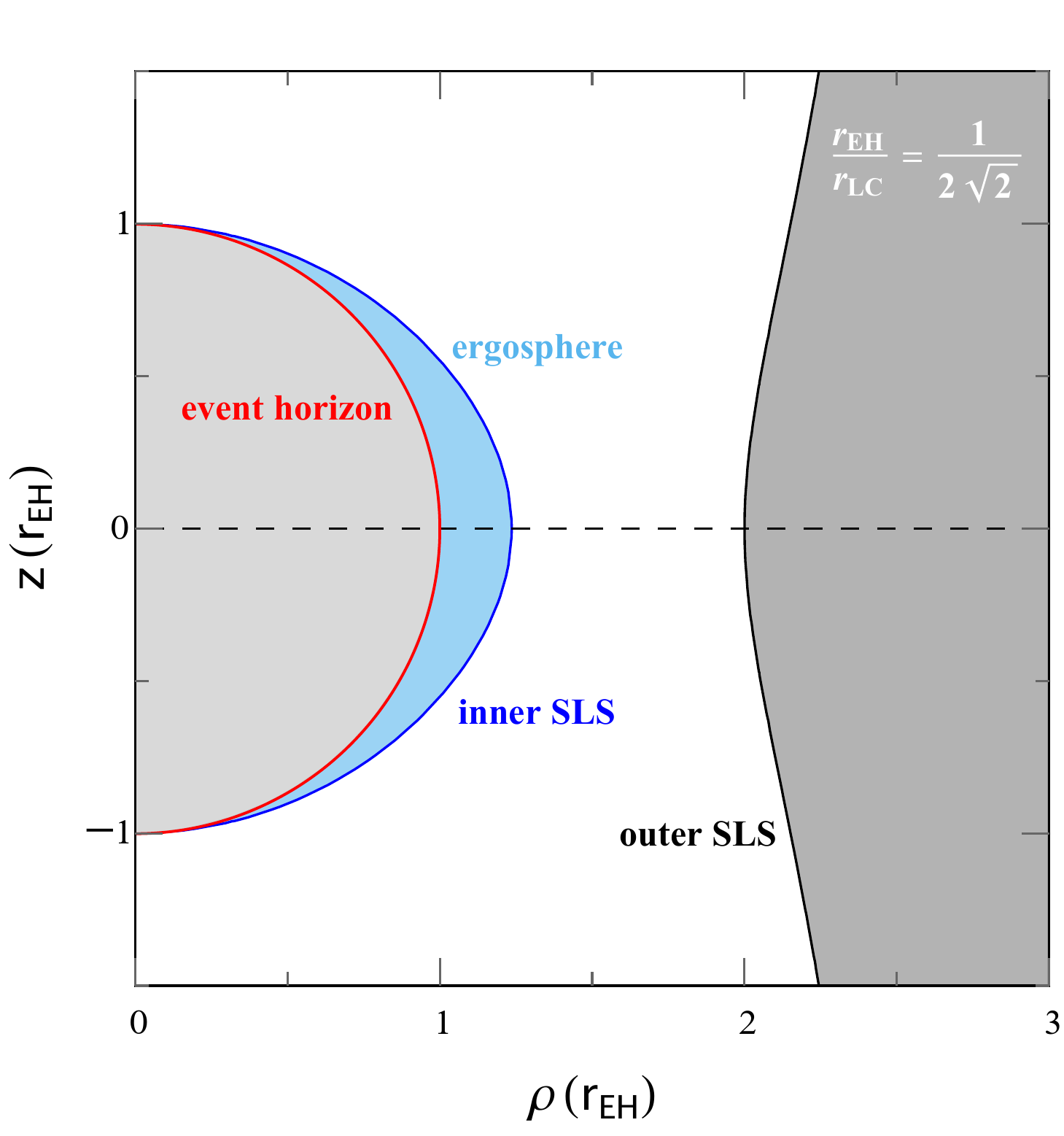}
}
\caption{
Horizon structure around the rotational metric solution with a singularity at the origin (side view). Here, the structure is presented for any given $\varphi$. The red curve marks the event horizon, whereas the light blue region represents the ergosphere. The inner and outer stationary limit surfaces (SLSs) are colored blue and black, respectively. The white region between the two SLSs corresponds to a rigidly-rotating normal space. For an NS, it is associated with the magnetospheric region. 
}
\label{fig:ergosphere}
\end{figure}

As Fig.~\ref{fig:ergosphere} illustrates, from inside out we have
a curvature singularity (origin), a horizon (red), an ergosphere (light blue), an inner SLS (blue), a normal space (white), and an outer SLS (black).
Similar to the McVittie metric~\citep{McVittie1933}, the rotational metric could evolve over time.
Compared with other well-known metric solutions, such as the Schwarzchild, Kerr, and Schwarzschild-de Sitter metrics~\citep{Kottler1918}, 
the rotational metric given by Eq.~\eqref{eq:RotSchM} exhibits a different horizon structure.
Nonetheless, the rotational metric $\mathrm{d} s_{\omega}^2$ can be easily compared with these solutions in geometry.
For instance, the solution behaves as that of Schwarzchild~\citep{Schwarzschild03} at the poles where the inner SLS is tangent to the event horizon~\citep{Weinberg:1972kfs,Rindler2006book,Carroll2014},
but also as that of Kerr~ \citep{Kerr1963} in the ergosphere where massive particles are necessarily dragged along with the BH's rotation, 
and finally like that of Schwarzschild-de Sitter (SdS) in the faraway region outside the outer SLS, where it is not asymptotically flat~\citep{Kottler1918,Zhang:2021ygh,Zhang:2023neo}.
Interestingly, there is still a normal space between the inner and outer SLSs, as represented in Fig.~\ref{fig:ergosphere} by the white region.
The normal space rigidly corotates with the spinning mass at an angular velocity $\omega$ and directly connects with the usual space surrounding us. 
For instance, it can be linked to a corotating NS magnetosphere, as presented below.

It is crucial to understand the physical origin of the ergosphere. We can describe this origin in terms of physical degrees of freedom. 
To be more specific, the rotational transformation given by Eq.~\eqref{eq:phi-dphi} contains a new dynamical quantity, i.e., $\omega=\omega(t)$, and it carries additional physical degrees of freedom.
However, these physical degrees of freedom are absorbed by the gravitational system during the transformation process.
It appears that the gravitational system has `eaten' these physical degrees of freedom through the non-inertial transformation \eqref{eq:phi-dphi},
leading to the formation of an additional ergosphere inside the system. 
This is somewhat analogous to the Higgs mechanism observed in particle physics \citep{Schwartz14}.
This notable phenomenon can also be described using a simple symmetry argument: 
the Schwarzschild metric processes more mutually independent Killing vectors than the rotational metric obtained through 
the non-inertial transformation given by Eq. \eqref{eq:phi-dphi}; in other words, the former exhibits greater symmetry than the latter, as detailed in Appendix~\ref{app:A}. 
Note that Killing vectors are defined in a coordinate- or frame-dependent manner, although the associated conserved quantities are coordinate-independent \citep{Weinberg:1972kfs,Rindler2006book,Carroll2014}.
It is evident that the symmetry has been reduced or partially broken by the non-inertial transformation (Eq. \eqref{eq:phi-dphi}). Thus, the non-inertial transformation plays a role of `symmetry breaking'. Hence, the rotational metric can be a compelling example that demonstrates the role of non-inertial transformations in introducing new structures into a gravitational system.

\section{Gravitational Effects} 
\label{sec:4} 

For a steadily rotating system, such as the magnetosphere of an NS, 
the angular-momentum parameter $\omega$ can be considered constant \citep{Goldreich96}, 
at least for a short time period spanning a few months or years, which is the case we focus on in this section.
In the case of NSs, $\omega$ represents the rate of rotation.
Next, we come back to the standard form of the rotational metric $\mathrm{d} s_{\omega}^2$. In general, the representation of any stationary metric is form-invariant under the following coordinate transformation \citep{Rindler2006book}:
\begin{equation}
\label{eq:TTransf}
\begin{array}{rcl}
\displaystyle
t\mapsto\,t^{\prime}=\kappa\left[t+\eta(x^{1}, x^{2}, x^{3})\right], 
\end{array}
\end{equation}
where $\kappa$ is a constant parameter and $\eta=\eta(x^{1}, x^{2}, x^{3})$ is a continuous function of the space-like coordinates $\{x^{i}, i=1,2,3\}$.
Here, time $t$ is changed while other coordinates such as $r$ and $\rho$ remain the same. 
Under the coordinate transformation given by Eq. \eqref{eq:TTransf}, $k_{ij}$ remains invariant \citep{Rindler2006book}, while $\psi$ and $\upsilon_{i}$ transform as
\begin{equation}
\label{eq:TTransf2}
\begin{array}{rcl}
\displaystyle
\psi\mapsto\psi^{\prime}=\psi-{\rm ln}\,\kappa, ~~ \upsilon_{i}\mapsto \upsilon_{i}^{\prime}=\kappa\left(\upsilon_{i}+\partial\,\!\eta/\partial\,\!x^{i}\right).
\end{array}
\end{equation}
These are the gauge coordinate transformations presented in \cite{Rindler2006book}. 
According to Eq. \eqref{eq:RotSchM}, one has
\begin{equation}
\label{eq:vecw}
\begin{array}{rcl}
\displaystyle
\vec{\upsilon}=\left(\upsilon_{1},\upsilon_{2},\upsilon_{3}\right)=\left(0,0, -\frac{\rho^2 \omega}{1\!-2\,\frac{\,M\,}{r}-\rho^2 \omega^2}\right),
\end{array}
\end{equation}
which can be transformed away. Indeed, there always exists an $\eta$ such that $\upsilon_{i}=-\partial\,\!\eta/\partial\,\!x^{i}$, which is known as the Rindler gauge  \citep{Rindler2006book}. From this point on, our analysis is carried out within the framework of the Rindler gauge.

In current experiments, the range of interest is set by $-\,g_{00}=e^{2\psi}>0$, indicating that $r^{-}_{\theta}< r< r^{+}_{\theta}$. For small values of $\omega$ and $M$, one has $r^{-}_{\theta}\simeq r_{\rm EH}$ and $r^{+}_{\theta}\simeq r_{{\rm LC}}$. 
In this range, according to Eq. \eqref{eq:RotSchM}, we obtain 
\begin{equation}
\label{eq:Spotential}
\begin{array}{rcl}
\displaystyle
\psi=\frac{1}{2}\,{\rm ln}\left(1\!-2\frac{\,M\,}{r}-\rho^2 \omega^2\right),
\end{array}
\end{equation}
which is the relativistic scalar potential. 
Under the weak field approximation, we have that
\begin{equation}
\label{eq:Spotential2}
\begin{array}{rcl}
\displaystyle
\psi\simeq-\frac{\,M\,}{r}-\frac{1}{2}\,\rho^2 \omega^2,
\end{array}
\end{equation}
where the outward `centrifugal' force can be generated by the $\omega$-term.
Note that, in GR, the centrifugal force is an effect of inertia-gravity (or gravity of inertia) \citep{Rindler2006book}.
In the Rindler gauge, the reference system is coordinate time synchronizable. Physically, coordinate time $t$ and proper time $\tau$ are connected by
\begin{equation}
\label{eq:clocks}
\begin{array}{rcl}
\displaystyle
\mathrm{d}\tau=e^{\psi}\,\mathrm{d} t =\mathrm{d} t \,\sqrt{1\!-2\,\frac{\,M\,}{r}-\rho^2 \omega^2},
\end{array}
\end{equation}
where $\tau$ can be measured by local observers. This has a clear physical meaning. 
In fact, the formula represents gravitational time dilation, providing an exact solution to Einstein’s clock-rate problem in curved spacetime.
Specifically, the time dilation reduces the familiar gravitational redshift observed in the (static) Schwarzschild system
as the angular-momentum parameter $\omega$ approaches zero.
Likewise, as the mass parameter $M$ approaches zero, it can replicate the time dilation effect observed in Einstein's uniformly rotating disk \citep{Rindler2006book}, 
which is sometimes referred to as the transverse Doppler effect in astronomy and astrophysics.
According to Eq.~\eqref{eq:clocks}, the gravitational redshift, when measured from a static observer, diverges as the observer approaches either SLS.
Therefore, in our case, both SLSs are infinite redshift surfaces, similar to existing solutions~\citep{Rindler2006book}. 

As shown by A. Einstein in his article on special relativity, time passes more slowly at the Earth's equator than at the poles \citep{Einstein1952}. 
The problem can be extended to curved spacetime.
In fact, within GR, the rotational metric provides a more comprehensive solution to the clock-rate problem in curved spacetime. Here, we denote ${\rm d}\tau_{e}$ as a proper separation at the equator and ${\rm d}\tau_{p}$ as that at the poles.
According to Eq.~\eqref{eq:clocks}, one has 
\begin{equation}
\label{eq:clockep0}
\begin{array}{rcl}
\displaystyle
\frac{{\rm d}\tau_{e}}{{\rm d}\tau_{p}}=\sqrt{\frac{1\!-2M/R- \omega^2R^2}{1\!-2M/R}}<1,
\end{array}
\end{equation}
where $R$ denotes the radius of an astrophysical object.
In the weak-field limit of $M/R\ll1$ and $\omega^2 R^2 \ll1$, 
\begin{equation}
\label{eq:clockep1}
\nonumber
{\rm d}\tau_{e}-{\rm d}\tau_{p}\simeq-\frac{1}{2}\,\omega^2 R^2{\rm d}\tau_{p},
\end{equation}
which further confirms Einstein's result.
In view of this, the rotating metric can be applied universally in everyday life. For instance, when comparing the passage of time between two locations on Earth, we must take into account not only the gravitational effects induced by Earth's mass but also factors such as latitude and rotation. Hence, the rotational metric is particularly relevant for technologies that require precise time measurement, such as the BeiDou Navigation Satellite System and the Global Positioning System.

\section{Magnetospheric Applications} 
\label{sec:5} 

Next, let us explore some potential applications of the rotational metric $\mathrm{d} s_{\omega}^2$ for describing certain astrophysical phenomena. 
For instance, consider an NS whose radius, $R_{\rm NS}\sim10$ km is approximately of the same order of magnitude as its Schwarzschild radius, $r_{\rm EH}$, 
but for which the gravitational system is free of any singularities and horizons.
Generally, the magnetospheric plasma is `frozen' to the magnetic field of the NS and 
therefore compelled to corotate steadily with the NS’s spin \citep{Goldreich96}.
Thus, the rotational metric may be applied to describe the rotationally-induced gravitational effects experienced by the plasma within the corotating magnetosphere. 
The corotating magnetospheric region extends from the NS surface at $r=R_{\rm NS}$ to the light cylinder at $r\sim r_{\rm LC}$.
As an example, the Galactic magnetar SGR J1935+2154$-$a highly magnetized NS$-$ has been measured to have a spin period of $P\approx$ 3.245 s \citep{Israel16}. 
If we assume a typical mass value for the magnetar, such as $M= 1.4~M_{\odot}$,
we can estimate that $r_{\rm EH}\sim4$ km and $r_{\rm LC}\sim1.6\times10^{5}$ km, where $M_{\odot}$ is the mass of the Sun. 
In this realistic scenario, $r_{\rm EH}/r_{\rm LC}\sim2.7\times10^{-5}$.
Then, we can approximate the lower boundary of the region in the normal space as $r^{-}_{\theta}\simeq\,r_{\rm EH}$ and the upper boundary as $r^{+}_{\theta}\simeq\,r_{\rm LC}$. In the physics of NSs, the magnetosphere is defined outwardly by the light cylinder.
Hence, the rotational metric~$\mathrm{d} s_{\omega}^2$ can describe the rotationally-induced gravitational effects (see Section \ref{sec:4}) 
experienced by the magnetospheric plasma, especially when $\rho\lesssim r_{\rm LC}$.
Beyond the light cylinder, i.e., when $\rho> r_{\rm LC}$, the plasma is unable to corotate with the NS, 
as doing so would result in its corotational velocity exceeding the speed of light.
Thus, the rotational metric is not applicable in the faraway region.
However, the Schwarzschild metric can describe the region far outside the light cylinder.
By the Einstein field equation, an alternative metric, the {\it magnetospheric metric}, is always available to describe the region around $\rho\simeq r_{\rm LC}$, 
which continuously connects with the rotational metric inwards and reduces to the Schwarzschild metric in the distant region (Appendix~\ref{app:A}).
Therefore, the rotational and Schwarzschild metrics need to be simultaneously used to describe different regions in one gravitational system,
which implies the necessity for the rotational metric to exist independently of the original Schwarzschild metric.

For comparison, we also examine the Kerr metric alongside the rotational metric.
Let us denote $a$, $I$, and $j$ as the angular momentum per unit mass, moment of inertia, and dimensionless spin parameter of the magnetar SGR J1935+2154, respectively.
Here, $j=a\,/M$.
For a typical NS, $I\sim\frac{2}{5} \,M R_{\mathrm{NS}}^{\,2}$.
Then, we can estimate the dimensionless parameter $j$ as 
\begin{equation}
\begin{array}{rcl}
\displaystyle
\nonumber
j=\frac{a}{M}=\frac{\omega I}{M^2}\sim\frac{1}{5}\,\frac{r_{\rm EH}}{r_{\rm LC}} \left(\frac{R_{\mathrm{NS}}}{M}\right)^{2}, 
\end{array}
\end{equation}
which is a basic parameter in the Kerr metric.
This is an approximation, but it is sufficient for an order-of-magnitude estimate.
Thus, one obtains $j\lesssim2.6\times{10^{-4}}$, which is $3-4$ orders of magnitude smaller than unity,
leading to the disappearance of the ergosphere in the Kerr solution.
Consequently, the Kerr metric reduces to the Schwarzschild metric. 
In the case of the rotational metric, $\Delta/r_{\rm EH}\sim10^{-9}$, causing the ergosphere to vanish in this scenario as well.
However, unlike the Kerr metric, our rotational metric features an axisymmetric normal space that rigidly rotates about a spin axis. 
It is important to note that this space is spatially infinite in the direction of the spin axis, making it an infinitely large space. 
In this normal space, the effects from the $\omega$-terms in the rotational metric remain substantial, 
given that the angular-momentum parameter $\omega$ does not change with the radius $r$. 
This contrasts with the Kerr case, where the angular velocity decreases significantly with the radius $r$ 
and can be approximated as ${\mathrm\Omega}\propto r^{-3}$ in the weak-field approximation \citep{GrHer2007book}.
As a result, the rotational metric can describe the rotationally-induced gravitational effects experienced by the magnetospheric plasma,
whereas the Kerr metric cannot.

\section{\hspace*{+0.0mm}Conclusions}
\label{sec:6}

In this study, using rotational transformations, 
we obtained a time-dependent metric solution to the Einstein field equations, namely, the rotational metric, for a rigidly-rotating volume of space.
In particular, it reduces to Einstein's metric for a uniformly rotating disk in the zero-mass limit.
We examined the effects of inertia-gravity arising from rotational transformations 
and demonstrated their influence on the metric structure, as well as the gravitational effects associated with rotation.
Interestingly, the rotational metric can be interpreted as a BH solution, featuring a curvature singularity, an event horizon, and an ergosphere. 
Notably, this solution can be derived from the Schwarzschild metric via a rotational transformation. 
The presence of the ergosphere can be attributed to the `non-inertial' nature of the transformation used.
We further explored the origin of the ergosphere by analyzing the physical degrees of freedom carried by the rotational transformation
and showed that this transformation partially breaks the symmetry associated with the original Schwarzschild solution prior to the rotational transformation.
In addition, we found that the rotational metric provides an exact solution to Einstein's clock-rate problem in curved spacetime,
and it accounts for various gravitational effects.
For example, a more general form of time-dilation emerges naturally from the solution, 
which reduces to the standard gravitational redshift and the transverse Doppler effect in the zero-rotation and zero-mass limits, respectively.
Finally, we observed that the metric possesses a rotating normal space between its two SLSs,
which can be directly associated with the corotating magnetospheres of NSs, normal stars, or other exotic astrophysical objects.

\section*{Acknowledgements}
We thank Profs.~Shu-Xu Yi, Shuang-Nan Zhang, and Shao-Lin Xiong for their useful discussions, along with the anonymous reviewers for their constructive comments and valuable suggestions. This work is supported by the National Program on Key Research and Development Project from the Ministry of Science and Technology of China (Grant No. 2021YFA0718500), as well as the funding from the Chinese Academy of Sciences (Grant Nos. E329A3M1 and E3545KU2), the Institute of High Energy Physics (Grant No. E25155U1), and the National Natural Science Foundation of China (Grant No. 12235001).

\section*{Data Availability} 
All data are available in the paper. 
\section*{Competing Interests} 
The authors declare that they have no conflicts of interest related to this work. \\


\appendix

\newpage
\onecolumngrid

\section{Metrics and Coordinate Transformations}	
\label{app:A}

\subsection{Magnetospheric Metric}
The rotational metric needs to exist independently of the Schwarzschild metric.
These two metrics can independently coexist to describe the regions within and beyond the NS magnetosphere, respectively. 
In the physics of NSs, the magnetospheric plasma corotates steadily with the NS's spin \citep{Goldreich96,1982RvMP...54....1M}. 
Generally, the corotational velocity of the plasma increases linearly with the cylindrical radius $\rho$ within the light cylinder.
When $\rho$ reaches or surpasses the light cylinder, the velocity can potentially exceed the speed of light.
Nevertheless, surpassing the speed of light is impossible for the plasma. 
To avoid this problem, the plasma can no longer be frozen to the magnetic field of the NS magnetosphere.
In reality, it may be no longer forced to corotate with the NS at a smaller radius than the light cylinder radius, i.e., at $\rho=r_{\rm CO}\doteq\left(1-\varepsilon\right)r_{\rm LC}$,
where $\varepsilon$ is an extremely small positive parameter. Hereafter, $r_{\rm CO}$ is referred to as the {\it corotational radius}, within which the plasma corotates with the spinning NS \citep{Goldreich96,1982RvMP...54....1M}. 
In general, $r_{\rm CO}\sim r_{\rm LC}$ is expected in physics of NSs. 
Consequently, in the region with $r\lesssim r_{\rm LC}$, the rotationally-induced gravitational effects experienced by the plasma can be expressed using the rotational metric $\mathrm{d} s_{\omega}^2$.
Simultaneously, when the plasma is far outside the light cylinder, 
it does not corotate with the NS. Therefore, the gravitational effects experienced by this plasma cannot be described using the rotational metric. 
Instead, these effects can be described using the Schwarzschild metric.
In principle, an additional metric can be constructed to describe the region with $\rho>r_{\rm CO}$ from the rotational metric $\mathrm{d} s_{\omega}^2$ by replacing $\omega$ with $\omega\,\Theta\left(\rho, z\right)$. Exactly, the components of this metric can be exactly expressed in cylindrical coordinates $\left(t, \rho, z, \varphi\right)$ as
\begin{equation}
\label{eq:A3rdRMetric}
g_{\mu\nu }\!=\!\!\left[\!
\begin{matrix}
-\!1\!+\!\frac{2 M}{r}\!+\!\left(\omega\,\Theta\right)^2 \rho ^2 \!\!&\!\! 0 \!\!&\!\!0 \!\!&\!\! -\left(\omega\,\Theta\right)\rho ^2 \\
 0 \!\!&\!\! 1\!+\!\frac{2 M \rho ^2}{r^{3} \left(1\!-\!\frac{2 M}{r}\right)} \!\!&\!\! \frac{2 M \rho  z}{ r^{3} \left(1\!-\!\frac{2 M}{r}\right)} \!\!&\!\! 0 \\
 0 \!\!&\!\! \frac{2 M \rho  z}{ r^{3} \left(1\!-\!\frac{2 M}{r}\right)} \!\!&\!\! 1\!+\!\frac{2 M z^2}{r^3 \left(1\!-\!\frac{2 M}{r}\right)} \!\!&\!\! 0 \\
-\left(\omega\,\Theta\right)\rho ^2  \!\!&\!\! 0 \!\!&\!\! 0 \!\!&\!\! \!\rho ^2 \\
\end{matrix}
\right],
\end{equation}
where the function $\Theta\equiv\Theta\left(\rho, z\right)$ is fully determined by the local environment, tends to decrease from 1 at $\rho=r_{\rm CO}$, 
and approaches zero as $r$ increases.
The exact form of the $\Theta$ function can be obtained by solving the Einstein field equations in the vicinity of $\rho\sim r_{\rm LC}$. 
Therefore, there always exists a metric that continuously connects the rotational metric $\mathrm{d} s_{\omega}^2$ inwards and the Schwarzschild metric outwards. 
The newly constructed metric \eqref{eq:A3rdRMetric} is designated as the {\it magnetospheric metric}.
Indeed, it reduces to the Schwarzschild metric in the region far outside the light cylinder.
As a result, these metrics are capable of describing different regions, enabling accurate 
characterization of the gravitational effects experienced by the plasma within and outside the light cylinder, respectively.
In particular, the rotational metric $\mathrm{d} s_{\omega}^2$ can effectively describe the region within the light cylinder,
whereas the Schwarzschild metric is suitable for describing the region located far outside the light cylinder.
Another point of interest is that the rotational metric is not asymptotically flat in any direction other than the rotation axis. 
Luckily, this occurs in the faraway region with $\rho>r_{\rm CO}$.
In this faraway region, the asymptotic non-flatness of the rotational metric may give rise to a series of difficulties in interpreting the metric physically,
despite successful analyses conducted on the asymptotic non-flat region \cite[e.g.,][]{Zhang:2021ygh,Zhang:2023neo}. 
However, the asymptotic non-flatness of the outer region with $\rho>r_{\rm CO}$ has no influence on the applications of the rotational metric to the magnetospheric region with $\rho< r_{\rm CO}\sim r_{\rm LC}$. In short, to properly describe the different regions in an NS system, it is crucial to simultaneously use the rotational and Schwarzschild metrics. Hence, both metrics should coexist independently in order to provide a comprehensive understanding of the NS system.

\subsection{Symmetries and Physical Degrees of Freedom}
Compared to the original Schwarzschild metric, the rotational metric has additional physical degrees of freedom.
Clearly, the rotational metric features an additional dynamical parameter, namely the angular velocity $\omega$. 
Analysis can be conducted based on the symmetries of metrics in terms of their Killing vectors.
In the general case of $\omega\neq0$, the rotational metric has at least two fewer mutually independent Killing vectors than the Schwarzschild metric. 
In GR, a metric determines a gravitational system.
Each Killing vector generates a symmetry that corresponds to a coordinate-independent conserved physical quantity, such as energy or angular momentum. This imposes a physical constraint on the corresponding gravitational system. 
As a result, the smaller the number of Killing vectors a spacetime possesses, the more the physical degrees of freedom it contains. Therefore, the gravitational system described by the rotational metric has more physical degrees of freedom than the Schwarzschild system.
Consequently, the two metrics cannot be considered physically equivalent when describing specific gravitational systems; 
this distinction arises from the presence of inertia-gravity.
Mathematically, the rotational metric can be obtained from the Schwarzschild metric through coordinate transformations. 
However, these transformations are non-inertial and involve changes in the physical degrees of freedom. 
As such, the specific gravitational physics is altered during the non-inertial transformations.
In fact, there are several other pairs of metrics that exhibit similar relationships \citep{Stephani.book.2003,Griffiths2009,Petel22}. 
For instance, the McVittie metric is given by \citep{McVittie1933} 
\begin{equation}
\label{eq:McVittie0}
\begin{array}{rcl}
\displaystyle
\mathrm{d} S^2 =-\frac{\left[1-\frac{M}{4\,a(\hat{t})\hat{r}}\right]^2}{\left[1+\frac{M}{4\,a(\hat{t})\hat{r}}\right]^2}\,\mathrm{d} \hat{t}^{\,2} +a(\hat{t})^2\left[1+\frac{M}{4\,a(\hat{t})\hat{r}}\right]^4\left[\mathrm{d}\hat{r}^{\,2} +\hat{r}^{\,2}\!\left(\mathrm{d} \theta^2\!+\sin^{2}\!\theta\,\mathrm{d}\varphi^2\right)\right],
\end{array}
\end{equation}
where $\hat{t}$, $\hat{r}$, $\theta$, and $\varphi$ are spacetime coordinates. Here, $a=a(\hat{t})$ is the expansion factor.
We consider a special scenario where the universe is driven by the cosmological constant $\Lambda$.
In this case, the expansion factor is expressed as $a(\hat{t})={\rm Exp} \left(\sqrt{\frac{\Lambda}{3}} \,\hat{t}\right)$.
Let us introduce a coordinate transformation, $r=\left[1+\frac{M}{4\,a(\hat{t})\hat{r}}\right]^2 a(\hat{t})\,\hat{r}$. 
With this transformation, the metric can be reexpressed as follows:
\begin{equation}
\label{eq:McVittie1}
\begin{array}{rcl}
\displaystyle
\mathrm{d} S^2 =-\!\left(1-\frac{M}{r}-\frac{\Lambda}{3} r^2\right)\,\mathrm{d} \hat{t}^{\,2} \!+\!\frac{\mathrm{d} r^{2}}{1-M/r} \!-\!\frac{2\sqrt{\frac{\Lambda}{3}}\,r}{\sqrt{1-M/r}}\mathrm{d} \hat{t}\,\mathrm{d} r+r^{2}\!\left(\mathrm{d} \theta^2\!+\!\sin^{2}\!\theta\,\mathrm{d}\varphi^2\right).
\end{array}
\end{equation}
Next, we replace $\hat{t}$ with $t=\hat{t}+F(r)$, where $F(r)$ is determined by the relation:
\begin{equation}
\label{eq:McVittie2}
\begin{array}{rcl}
\displaystyle
\frac{\mathrm{d} F}{\mathrm{d} r}=\frac{\sqrt{\frac{\Lambda}{3}}\,r}{\left(1-\frac{M}{r}-\frac{\Lambda}{3} r^2\right)\sqrt{1-\frac{M}{r}}}.
\end{array}
\end{equation}
From this substitution, we obtain the Schwarzschild-de Sitter (SdS) metric,
\begin{equation}
\label{eq:McVittie3}
\begin{array}{rcl}
\displaystyle
\mathrm{d} S^2 =- \left(1\!-\frac{\,M\,}{r}-\frac{\Lambda}{3} r^2\!\right)\, \mathrm{d} t^{2}
+\frac{1}{\, \left(1\!-\frac{\,M\,}{r}-\frac{\Lambda}{3} r^2\!\right) \,}\mathrm{d} r^2\!+\! r^{2}\!\left(\mathrm{d} \theta^2\!+\sin^{2}\!\theta\,\mathrm{d}\varphi^2\right).
\end{array}
\end{equation}
It is worth noting that the expansion factor of the universe is not limited to the specific form adopted above.
In fact, the choice of expansion factor significantly influences the physical interpretation of a metric.
From this perspective, one could argue that the McVittie metric, derived using an arbitrary expansion factor, encapsulates richer physics than the Schwarzschild–de Sitter (SdS) metric, which is based on a specific expansion factor.
This analysis can also be approached by considering the physical degrees of freedom involved.
Notably, the McVittie metric possesses one less Killing vector than the SdS metric, meaning it has more physical degrees of freedom.
As a result, the McVittie metric is not physically equivalent to the SdS metric when describing relevant gravitational systems.
Therefore, even if two metrics can be transformed into each other through coordinate transformations, this does not necessarily imply physical equivalence in representing specific gravitational systems or states, 
unless they share the same physical degrees of freedom and there is no change in `gravity of inertia' or `gravity of gravity' \citep{Rindler2006book}.
In general, after applying non-inertial transformations along with the standard transformations given by Eqs. \eqref{eq:TTransf} and  \eqref{eq:TTransf2},  
stationary metrics can be rewritten in the form
\begin{equation}
\label{eq:MMetric}
\begin{array}{rcl}
\displaystyle
\mathrm{d} S^2 =- \mathrm{d} T^{2}+\mathrm{d} X^{2}+\mathrm{d} Y^{2}+\mathrm{d} Z^{2},
\end{array}
\end{equation}
where $T$, $X$, $Y$, and $Z$ are the time and space coordinates in a locally inertial frame, respectively. 
It is evident that these metrics are not physically equivalent to the Minkowski metric, although they can be transformed into the latter under coordinate transformations.

\section{Calculations}
\label{app:B}

In this study, the rotational metric $\mathrm{d} s_{\omega}^2$ of the form given by Eq.~\eqref{eq:RotSchM} has a Lorentzian signature ${\rm diag}\left(-,+,+,+\right)$. It is also continuous and nondegenerate.
Generally, the Christoffel symbols are defined by
\begin{equation}
\label{eq:Christoffel}
\Gamma^{\lambda}_{\,\,\,\mu\nu}=\frac{1}{2}\,g^{\lambda\rho}\left(\partial_{\mu}\,g_{\rho\nu}+\partial_{\nu}\,g_{\rho\mu}-\partial_{\rho}\,g_{\mu\nu}\right),
\end{equation}
with $\partial_{\mu}=\partial /\partial x^{\mu}$. In the cylindrical coordinates, we have $x^{\mu}=\left(t, \rho,z,\varphi\right)$.

\noindent \\
From the metric $\mathrm{d} s_{\omega}^2$ and its inverse with $\omega=\omega(t)$, we obtain the nonzero Christoffel symbols: 
\begin{eqnarray}
\label{eq:ChristoffelSs}
\nonumber
\Gamma ^{\mathrm{t}}{}_{\mathrm{t}}{}_{\rho }&=&\Gamma ^{\mathrm{t}}{}_{\rho}{}_{ \mathrm{t}}=\frac{G M \rho }{\left(\rho ^2+z^2\right) \left(c^2 \sqrt{\rho ^2+z^2}-2 G M\right)}, \nonumber\\
\nonumber
\Gamma ^{\mathrm{t}}{}_{\mathrm{t}}{}_{\mathrm{z}}&=&\Gamma ^{\mathrm{t}}{}_{\mathrm{z}}{}_{\mathrm{t}}=\frac{G M z}{\left(\rho ^2+z^2\right) \left(c^2 \sqrt{\rho ^2+z^2}-2 G M\right)}, \nonumber\\
\nonumber
\Gamma ^{\rho }{}_{\mathrm{t}}{}_{\mathrm{t}}&=&\frac{\rho}{c^4} \left(\frac{G M \left(c^2+2 \rho ^2 \omega^2\right)}{\left(\rho ^2+z^2\right)^{3/2}}-c^2 \omega^2-\frac{2 G^2 M^2}{\left(\rho ^2+z^2\right)^2}\right), \nonumber\\
\nonumber
\Gamma ^{\rho }{}_{\mathrm{t}}{}_{\varphi }&=&\Gamma ^{\rho }{}_{\varphi }{}_{\mathrm{t}}=\frac{\rho  \omega}{c^3} \left(c^2-\frac{2 G M \rho ^2}{\left(\rho ^2+z^2\right)^{3/2}}\right), \nonumber\\
\nonumber
\Gamma ^{\rho }{}_{\rho }{}_{\rho }&=&\frac{G M \rho  \left(c^4 \left(-\rho ^4+2 z^4+\rho ^2 z^2\right)+2 c^2 G M \left(\rho ^2-4 z^2\right) \sqrt{\rho ^2+z^2}+8 G^2 M^2 z^2\right)}{\left(\rho ^2+z^2\right)^{5/2} \left(c^3 \sqrt{\rho ^2+z^2}-2 c G M\right)^2}, \nonumber\\
\nonumber
\Gamma ^{\rho }{}_{\rho }{}_{\mathrm{z}}&=&\Gamma ^{\rho }{}_{\mathrm{z}}{}_{\rho }=-\frac{G M \rho ^2 z \left(3 c^4 \left(\rho ^2+z^2\right)+2 G M \left(4 G M-5 c^2 \sqrt{\rho ^2+z^2}\right)\right)}{\left(\rho ^2+z^2\right)^{5/2} \left(c^3 \sqrt{\rho ^2+z^2}-2 c G M\right)^2},\nonumber\\
\nonumber
\Gamma ^{\rho }{}_{\mathrm{z}}{}_{\mathrm{z}}&=&\frac{G M \rho  \left(c^4 \left(2 \rho ^4-z^4+\rho ^2 z^2\right)+2 c^2 G M \left(z^2-4 \rho ^2\right) \sqrt{\rho ^2+z^2}+8 G^2 M^2 \rho ^2\right)}{\left(\rho ^2+z^2\right)^{5/2} \left(c^3 \sqrt{\rho ^2+z^2}-2 c G M\right)^2},\nonumber\\
\nonumber
\Gamma ^{\rho }{}_{\varphi }{}_{\varphi }&=&\frac{2 G M \rho ^3}{c^2 \left(\rho ^2+z^2\right)^{3/2}}-\rho,\\
\nonumber
\Gamma ^{\mathrm{z}}{}_{\mathrm{t}}{}_{\mathrm{t}}&=&\frac{G M z \left(\sqrt{\rho ^2+z^2} \left(c^2+2 \rho ^2 \omega^2\right)-2 G M\right)}{c^4 \left(\rho ^2+z^2\right)^2},\\
\nonumber
\Gamma ^{\mathrm{z}}{}_{\mathrm{t}}{}_{\varphi }&=&\Gamma ^{\mathrm{z}}{}_{\varphi }{}_{\mathrm{t}}=-\frac{2 G M \rho ^2 \omega z}{c^3 \left(\rho ^2+z^2\right)^{3/2}},\\
\nonumber
\Gamma ^{\mathrm{z}}{}_{\rho }{}_{\rho }&=&\frac{G M z \left(c^4 \left(-\rho ^4+2 z^4+\rho ^2 z^2\right)+2 c^2 G M \left(\rho ^2-4 z^2\right) \sqrt{\rho ^2+z^2}+8 G^2 M^2 z^2\right)}{\left(\rho ^2+z^2\right)^{5/2} \left(c^3 \sqrt{\rho ^2+z^2}-2 c G M\right)^2},\\
\nonumber
\Gamma ^{\mathrm{z}}{}_{\rho }{}_{\mathrm{z}}&=&\Gamma ^{\mathrm{z}}{}_{\rho }{}_{\mathrm{z}}=-\frac{G M \rho  z^2 \left(3 c^4 \left(\rho ^2+z^2\right)+2 G M \left(4 G M-5 c^2 \sqrt{\rho ^2+z^2}\right)\right)}{\left(\rho ^2+z^2\right)^{5/2} \left(c^3 \sqrt{\rho ^2+z^2}-2 c G M\right)^2},\\
\nonumber
\Gamma ^{\mathrm{z}}{}_{\mathrm{z}}{}_{\mathrm{z}}&=&\frac{G M z \left(c^4 \left(2 \rho ^4-z^4+\rho ^2 z^2\right)+2 c^2 G M \left(z^2-4 \rho ^2\right) \sqrt{\rho ^2+z^2}+8 G^2 M^2 \rho ^2\right)}{\left(\rho ^2+z^2\right)^{5/2} \left(c^3 \sqrt{\rho ^2+z^2}-2 c G M\right)^2},\\
\nonumber
\Gamma ^{\mathrm{z}}{}_{\varphi }{}_{\varphi }&=&\frac{2 G M \rho ^2 z}{c^2 \left(\rho ^2+z^2\right)^{3/2}},\\
\nonumber
\Gamma ^{\varphi }{}_{\text{t}}{}_{\text{t}}&=&-\frac{1}{c^2}\frac{\partial \omega}{\partial t},\\
\nonumber
\Gamma ^{\varphi }{}_{\mathrm{t}}{}_{\rho }&=&\Gamma ^{\varphi }{}_{\rho }{}_{\mathrm{t}}=\frac{\omega \left(G M \left(3 \rho ^2+2 z^2\right)-c^2 \left(\rho ^2+z^2\right)^{3/2}\right)}{c \rho  \left(\rho ^2+z^2\right) \left(c^2 \sqrt{\rho ^2+z^2}-2 G M\right)},\\
\nonumber
\Gamma ^{\varphi }{}_{\mathrm{t}}{}_{\mathrm{z}}&=&\Gamma ^{\varphi }{}_{\mathrm{z}}{}_{\mathrm{t}}=\frac{G M \omega z}{c \left(\rho ^2+z^2\right) \left(c^2 \sqrt{\rho ^2+z^2}-2 G M\right)},\\
\nonumber
\Gamma ^{\varphi }{}_{\varphi }{}_{\rho }&=&\Gamma ^{\varphi }{}_{\rho }{}_{\varphi }=\frac{1}{\rho }.
\end{eqnarray}

\noindent \\
Recall that the Riemann tensor is defined in general as
\begin{equation}
\label{eq:Riemanntensor} 
R^{\lambda }{}_{\mu \nu \sigma }=-\frac{\partial \Gamma ^{\lambda }{}_{\mu \nu }}{\partial x^{\sigma} }+\frac{\partial \Gamma ^{\lambda }{}_{\mu \sigma }}{\partial x^{\nu} }+\Gamma ^{\eta }{}_{\mu \sigma } \Gamma ^{\lambda }{}_{\eta \nu }-\Gamma ^{\eta }{}_{\mu \nu } \Gamma ^{\lambda }{}_{\eta \sigma }.
\end{equation}
Substituting the Christoffel symbols into this formula, we obtain the nonvanishing components of the Riemann tensor:
\begin{eqnarray}
\label{eq:RiemannTs1}
\nonumber
\mathrm{R}^{\mathrm{t}}{}_{\mathrm{t}}{}_{\mathrm{t}}{}_{\varphi }&=&-\,\mathrm{R}^{\mathrm{t}}{}_{\mathrm{t}}{}_{\varphi }{}_{\mathrm{t}}=\frac{G M \rho ^2 \omega}{c^3 \left(\rho ^2+z^2\right)^{3/2}},\\
\nonumber
\mathrm{R}^{\mathrm{t}}{}_{\rho }{}_{\mathrm{t}}{}_{\rho }&=&-\,\mathrm{R}^{\mathrm{t}}{}_{\rho }{}_{\rho }{}_{\mathrm{t}}=\frac{G M \left(c^4 \left(2 \rho ^4-z^4+\rho ^2 z^2\right)+4 c^2 G M \rho ^2 \sqrt{\rho ^2+z^2}+4 G^2 M^2 z^2\right)}{c^2 \left(\rho ^2+z^2\right)^{5/2} \left(c^4 \left(\rho ^2+z^2\right)-4 G^2 M^2\right)},\\
\nonumber
\mathrm{R}^{\mathrm{t}}{}_{\rho }{}_{\mathrm{t}}{}_{\mathrm{z}}&=&-\,\mathrm{R}^{\mathrm{t}}{}_{\rho }{}_{\mathrm{z}}{}_{\mathrm{t}}=\frac{G M \rho  z \left(3 c^4 \left(\rho ^2+z^2\right)+4 G M \left(c^2 \sqrt{\rho ^2+z^2}-G M\right)\right)}{c^2 \left(\rho ^2+z^2\right)^{5/2} \left(c^4 \left(\rho ^2+z^2\right)-4 G^2 M^2\right)},\\
\nonumber
\mathrm{R}^{\mathrm{t}}{}_{\mathrm{z}}{}_{\mathrm{t}}{}_{\rho }&=&-\,\mathrm{R}^{\mathrm{t}}{}_{\mathrm{z}}{}_{\rho }{}_{\mathrm{t}}=\frac{G M \rho  z \left(3 c^4 \left(\rho ^2+z^2\right)+4 G M \left(c^2 \sqrt{\rho ^2+z^2}-G M\right)\right)}{c^2 \left(\rho ^2+z^2\right)^{5/2} \left(c^4 \left(\rho ^2+z^2\right)-4 G^2 M^2\right)},\\
\nonumber
\mathrm{R}^{\mathrm{t}}{}_{\mathrm{z}}{}_{\mathrm{t}}{}_{\mathrm{z}}&=&-\,\mathrm{R}^{\mathrm{t}}{}_{\mathrm{z}}{}_{\mathrm{z}}{}_{\mathrm{t}}=\frac{G M \left(c^4 \left(-\rho ^4+2 z^4+\rho ^2 z^2\right)+4 c^2 G M z^2 \sqrt{\rho ^2+z^2}+4 G^2 M^2 \rho ^2\right)}{c^2 \left(\rho ^2+z^2\right)^{5/2} \left(c^4 \left(\rho ^2+z^2\right)-4 G^2 M^2\right)},\\
\nonumber
\mathrm{R}^{\mathrm{t}}{}_{\varphi }{}_{\varphi }{}_{\mathrm{t}}&=&-\,\mathrm{R}^{\mathrm{t}}{}_{\varphi }{}_{\mathrm{t}}{}_{\varphi }=\frac{G M \rho ^2}{c^2 \left(\rho ^2+z^2\right)^{3/2}},\\
\nonumber
\mathrm{R}^{\rho }{}_{\mathrm{t}}{}_{\mathrm{t}}{}_{\rho }&=&-\,\mathrm{R}^{\rho }{}_{\mathrm{t}}{}_{\rho }{}_{\mathrm{t}}=\frac{G M \left(c^2 \left(2 \rho ^4-z^4+\rho ^2 z^2\right)+2 G M \left(z^2-2 \rho ^2\right) \sqrt{\rho ^2+z^2}+\omega^2 \left(\rho ^6-2 \rho ^2 z^4-\rho ^4 z^2\right)\right)}{c^4 \left(\rho ^2+z^2\right)^{7/2}},\\%
\nonumber
\mathrm{R}^{\rho }{}_{\mathrm{t}}{}_{\mathrm{t}}{}_{\mathrm{z}}&=&-\,\mathrm{R}^{\rho }{}_{\mathrm{t}}{}_{\mathrm{z}}{}_{\mathrm{t}}=\frac{3 G M \rho  \left(z \left(\rho ^2+z^2\right) \left(c^2+\rho ^2 \omega^2\right)-2 G M z \sqrt{\rho ^2+z^2}\right)}{c^4 \left(\rho ^2+z^2\right)^{7/2}},\\
\nonumber
\mathrm{R}^{\rho }{}_{\mathrm{t}}{}_{\rho }{}_{\varphi }&=&-\,\mathrm{R}^{\rho }{}_{\mathrm{t}}{}_{\varphi }{}_{\rho }=\frac{G M \rho ^2 \omega \left(\rho ^2-2 z^2\right)}{c^3 \left(\rho ^2+z^2\right)^{5/2}},\\
\nonumber
\mathrm{R}^{\rho }{}_{\mathrm{t}}{}_{\mathrm{z}}{}_{\varphi }&=&-\,\mathrm{R}^{\rho }{}_{\mathrm{t}}{}_{\varphi }{}_{\mathrm{z}}=\frac{3 G M \rho ^3 \omega z}{c^3 \left(\rho ^2+z^2\right)^{5/2}},\\
\nonumber
\mathrm{R}^{\rho }{}_{\rho }{}_{\rho }{}_{\mathrm{z}}&=&-\,\mathrm{R}^{\rho }{}_{\rho }{}_{\mathrm{z}}{}_{\rho }=-\frac{2 G^2 M^2 \rho  z \left(c^2 \sqrt{\rho ^2+z^2}+2 G M\right)}{c^2 \left(\rho ^2+z^2\right)^{5/2} \left(c^4 \left(\rho ^2+z^2\right)-4 G^2 M^2\right)},\\
\nonumber
\mathrm{R}^{\rho }{}_{\mathrm{z}}{}_{\rho }{}_{\mathrm{z}}&=&-\,\mathrm{R}^{\rho }{}_{\mathrm{z}}{}_{\mathrm{z}}{}_{\rho }=-\frac{G M \left(c^4 \left(\rho ^2+z^2\right)^2+2 c^2 G M z^2 \sqrt{\rho ^2+z^2}-4 G^2 M^2 \rho ^2\right)}{c^2 \left(\rho ^2+z^2\right)^{5/2} \left(c^4 \left(\rho ^2+z^2\right)-4 G^2 M^2\right)},\\
\nonumber
\mathrm{R}^{\rho }{}_{\varphi }{}_{\mathrm{t}}{}_{\rho }&=&-\,\mathrm{R}^{\rho }{}_{\varphi }{}_{\rho }{}_{\mathrm{t}}=-\frac{G M \rho ^2 \omega \left(\rho ^2-2 z^2\right)}{c^3 \left(\rho ^2+z^2\right)^{5/2}},\\
\nonumber
\mathrm{R}^{\rho }{}_{\varphi }{}_{\mathrm{t}}{}_{\mathrm{z}}&=&-\,\mathrm{R}^{\rho }{}_{\varphi }{}_{\mathrm{z}}{}_{\mathrm{t}}=-\frac{3 G M \rho ^3 \omega z}{c^3 \left(\rho ^2+z^2\right)^{5/2}},\\
\nonumber
\mathrm{R}^{\rho }{}_{\varphi }{}_{\rho }{}_{\varphi }&=&-\,\mathrm{R}^{\rho }{}_{\varphi }{}_{\varphi }{}_{\rho }=-\frac{G M \rho ^2 \left(\rho ^2-2 z^2\right)}{c^2 \left(\rho ^2+z^2\right)^{5/2}},\\
\nonumber
\mathrm{R}^{\rho }{}_{\varphi }{}_{\mathrm{z}}{}_{\varphi }&=&-\,\mathrm{R}^{\rho }{}_{\varphi }{}_{\varphi }{}_{\mathrm{z}}=-\frac{3 G M \rho ^3 z}{c^2 \left(\rho ^2+z^2\right)^{5/2}},\\
\nonumber
\mathrm{R}^{\mathrm{z}}{}_{\mathrm{t}}{}_{\mathrm{t}}{}_{\rho }&=&-\,\mathrm{R}^{\mathrm{z}}{}_{\mathrm{t}}{}_{\rho }{}_{\mathrm{t}}=\frac{3 G M \rho  \left(z \left(\rho ^2+z^2\right) \left(c^2+\rho ^2 \omega^2\right)-2 G M z \sqrt{\rho ^2+z^2}\right)}{c^4 \left(\rho ^2+z^2\right)^{7/2}},\\
\nonumber
\mathrm{R}^{\mathrm{z}}{}_{\mathrm{t}}{}_{\mathrm{t}}{}_{\mathrm{z}}&=&-\,\mathrm{R}^{\mathrm{z}}{}_{\mathrm{t}}{}_{\mathrm{z}}{}_{\mathrm{t}}
=\frac{G M \left(c^2 \left(-\rho ^4+2 z^4+\rho ^2 z^2\right)+2 G M \left(\rho ^2-2 z^2\right) \sqrt{\rho ^2+z^2}\right)}{c^4 \left(\rho ^2+z^2\right)^{7/2}}\\%
\nonumber
&&~~~~~~~~~~+\frac{G M \rho ^2 \omega^2 \left(\rho ^2+z^2\right) \left(z^2-2 \rho ^2\right)}{c^4 \left(\rho ^2+z^2\right)^{7/2}},\\%
\nonumber
\mathrm{R}^{\mathrm{z}}{}_{\mathrm{t}}{}_{\rho }{}_{\varphi }&=&-\,\mathrm{R}^{\mathrm{z}}{}_{\mathrm{t}}{}_{\rho }{}_{\varphi }=\frac{3 G M \rho ^3 \omega z}{c^3 \left(\rho ^2+z^2\right)^{5/2}},\\
\nonumber
\mathrm{R}^{\mathrm{z}}{}_{\mathrm{t}}{}_{\mathrm{z}}{}_{\varphi }&=&-\,\mathrm{R}^{\mathrm{z}}{}_{\mathrm{t}}{}_{\varphi }{}_{\mathrm{z}}=\frac{G M \rho ^2 \omega \left(z^2-2 \rho ^2\right)}{c^3 \left(\rho ^2+z^2\right)^{5/2}},\\
\nonumber
\mathrm{R}^{\mathrm{z}}{}_{\rho }{}_{\rho }{}_{\mathrm{z}}&=&-\,\mathrm{R}^{\mathrm{z}}{}_{\rho }{}_{\mathrm{z}}{}_{\rho }=\frac{G M \left(c^4 \left(\rho ^2+z^2\right)^2+2 c^2 G M \rho ^2 \sqrt{\rho ^2+z^2}-4 G^2 M^2 z^2\right)}{c^2 \left(\rho ^2+z^2\right)^{5/2} \left(c^4 \left(\rho ^2+z^2\right)-4 G^2 M^2\right)},\\
\nonumber
\mathrm{R}^{\mathrm{z}}{}_{\mathrm{z}}{}_{\mathrm{z}}{}_{\rho }&=&-\,\mathrm{R}^{\mathrm{z}}{}_{\mathrm{z}}{}_{\rho }{}_{\mathrm{z}}=-\frac{2 G^2 M^2 \rho  z \left(c^2 \sqrt{\rho ^2+z^2}+2 G M\right)}{c^2 \left(\rho ^2+z^2\right)^{5/2} \left(c^4 \left(\rho ^2+z^2\right)-4 G^2 M^2\right)},\\
\nonumber
\mathrm{R}^{\mathrm{z}}{}_{\varphi }{}_{\mathrm{t}}{}_{\rho }&=&-\,\mathrm{R}^{\mathrm{z}}{}_{\varphi }{}_{\rho }{}_{\mathrm{t}}=-\frac{3 G M \rho ^3 \omega z}{c^3 \left(\rho ^2+z^2\right)^{5/2}},\\
\nonumber
\mathrm{R}^{\mathrm{z}}{}_{\varphi }{}_{\mathrm{t}}{}_{\mathrm{z}}&=&-\,\mathrm{R}^{\mathrm{z}}{}_{\varphi }{}_{\mathrm{z}}{}_{\mathrm{t}}=\frac{G M \rho ^2 \omega \left(2 \rho ^2-z^2\right)}{c^3 \left(\rho ^2+z^2\right)^{5/2}},\\
\nonumber
\mathrm{R}^{\mathrm{z}}{}_{\varphi }{}_{\rho }{}_{\varphi }&=&-\,\mathrm{R}^{\mathrm{z}}{}_{\varphi }{}_{\varphi }{}_{\rho }=-\frac{3 G M \rho ^3 z}{c^2 \left(\rho ^2+z^2\right)^{5/2}},\\
\nonumber
\mathrm{R}^{\mathrm{z}}{}_{\varphi }{}_{\mathrm{z}}{}_{\varphi }&=&-\,\mathrm{R}^{\mathrm{z}}{}_{\varphi }{}_{\varphi }{}_{\mathrm{z}}=\frac{G M \rho ^2 \left(2 \rho ^2-z^2\right)}{c^2 \left(\rho ^2+z^2\right)^{5/2}},\\
\nonumber
\mathrm{R}^{\varphi }{}_{\mathrm{t}}{}_{\mathrm{t}}{}_{\varphi }&=&-\,\mathrm{R}^{\varphi }{}_{\mathrm{t}}{}_{\varphi }{}_{\mathrm{t}}=\frac{G M \left(\frac{\rho ^2 \omega^2-c^2}{\left(\rho ^2+z^2\right)^{3/2}}+\frac{2 G M}{\left(\rho ^2+z^2\right)^2}\right)}{c^4},\\
\nonumber
\mathrm{R}^{\varphi }{}_{\rho }{}_{\mathrm{t}}{}_{\rho }&=&-\,\mathrm{R}^{\varphi }{}_{\rho }{}_{\rho }{}_{\mathrm{t}}=\frac{3 G M \omega \left(c^4 \left(\rho ^4-z^4\right)+2 c^2 G M \rho ^2 \sqrt{\rho ^2+z^2}+4 G^2 M^2 z^2\right)}{c^3 \left(\rho ^2+z^2\right)^{5/2} \left(c^4 \left(\rho ^2+z^2\right)-4 G^2 M^2\right)},\\
\nonumber
\mathrm{R}^{\varphi }{}_{\rho }{}_{\mathrm{t}}{}_{\mathrm{z}}&=&-\,\mathrm{R}^{\varphi }{}_{\rho }{}_{\mathrm{z}}{}_{\mathrm{t}}=\frac{6 G M \rho  \omega z \left(c^4 \left(\rho ^2+z^2\right)+G M \left(c^2 \sqrt{\rho ^2+z^2}-2 G M\right)\right)}{c^3 \left(\rho ^2+z^2\right)^{5/2} \left(c^4 \left(\rho ^2+z^2\right)-4 G^2 M^2\right)},\\
\nonumber
\mathrm{R}^{\varphi }{}_{\rho }{}_{\rho }{}_{\varphi }&=&-\,\mathrm{R}^{\varphi }{}_{\rho }{}_{\varphi }{}_{\rho }=-\frac{G M \left(c^4 \left(-\rho ^4+2 z^4+\rho ^2 z^2\right)+2 c^2 G M \left(\rho ^2-4 z^2\right) \sqrt{\rho ^2+z^2}+8 G^2 M^2 z^2\right)}{\left(\rho ^2+z^2\right)^{5/2} \left(c^3 \sqrt{\rho ^2+z^2}-2 c G M\right)^2},\\
\nonumber
\mathrm{R}^{\varphi }{}_{\rho }{}_{\mathrm{z}}{}_{\varphi }&=&-\,\mathrm{R}^{\varphi }{}_{\rho }{}_{\varphi }{}_{\mathrm{z}}=\frac{G M \rho  z \left(3 c^4 \left(\rho ^2+z^2\right)+2 G M \left(4 G M-5 c^2 \sqrt{\rho ^2+z^2}\right)\right)}{\left(\rho ^2+z^2\right)^{5/2} \left(c^3 \sqrt{\rho ^2+z^2}-2 c G M\right)^2},\\
\nonumber
\mathrm{R}^{\varphi }{}_{\mathrm{z}}{}_{\mathrm{t}}{}_{\rho }&=&-\,\mathrm{R}^{\varphi }{}_{\mathrm{z}}{}_{\rho }{}_{\mathrm{t}}=\frac{6 G M \rho  \omega z \left(c^4 \left(\rho ^2+z^2\right)+G M \left(c^2 \sqrt{\rho ^2+z^2}-2 G M\right)\right)}{c^3 \left(\rho ^2+z^2\right)^{5/2} \left(c^4 \left(\rho ^2+z^2\right)-4 G^2 M^2\right)},\\
\nonumber
\mathrm{R}^{\varphi }{}_{\mathrm{z}}{}_{\mathrm{t}}{}_{\mathrm{z}}&=&-\,\mathrm{R}^{\varphi }{}_{\mathrm{z}}{}_{\mathrm{z}}{}_{\mathrm{t}}=\frac{3 G M \omega \left(c^4 \left(z^4-\rho ^4\right)+2 c^2 G M z^2 \sqrt{\rho ^2+z^2}+4 G^2 M^2 \rho ^2\right)}{c^3 \left(\rho ^2+z^2\right)^{5/2} \left(c^4 \left(\rho ^2+z^2\right)-4 G^2 M^2\right)},\\
\nonumber
\mathrm{R}^{\varphi }{}_{\mathrm{z}}{}_{\rho }{}_{\varphi }&=&-\,\mathrm{R}^{\varphi }{}_{\mathrm{z}}{}_{\varphi }{}_{\rho }=\frac{G M \rho  z \left(3 c^4 \left(\rho ^2+z^2\right)+2 G M \left(4 G M-5 c^2 \sqrt{\rho ^2+z^2}\right)\right)}{\left(\rho ^2+z^2\right)^{5/2} \left(c^3 \sqrt{\rho ^2+z^2}-2 c G M\right)^2},\\
\nonumber
\mathrm{R}^{\varphi }{}_{\mathrm{z}}{}_{\mathrm{z}}{}_{\varphi }&=&-\,\mathrm{R}^{\varphi }{}_{\mathrm{z}}{}_{\varphi }{}_{\mathrm{z}}=-\frac{G M \left(c^4 \left(2 \rho ^4-z^4+\rho ^2 z^2\right)+2 c^2 G M \left(z^2-4 \rho ^2\right) \sqrt{\rho ^2+z^2}+8 G^2 M^2 \rho ^2\right)}{\left(\rho ^2+z^2\right)^{5/2} \left(c^3 \sqrt{\rho ^2+z^2}-2 c G M\right)^2},\\
\nonumber
\mathrm{R}^{\varphi }{}_{\varphi }{}_{\mathrm{t}}{}_{\varphi }&=&-\,\mathrm{R}^{\varphi }{}_{\varphi }{}_{\varphi }{}_{\mathrm{t}}=-\frac{G M \rho ^2 \omega}{c^3 \left(\rho ^2+z^2\right)^{3/2}}.
\end{eqnarray}

\noindent \\
The final step is to calculate the Ricci tensor according to its definition, namely $R_{\mu\nu}=R^{\varrho }{}_{\mu \varrho \nu}$.
From those components of the Riemann tensor, we find $R_{\mu\nu}=0$.
This demonstrates that the rotational metric $\mathrm{d} s_{\omega}^2$ is indeed an exact vacuum solution of the Einstein field equations.

\end{document}